\documentclass[a4paper,11pt]{article}
\usepackage{pos}
\usepackage{graphicx} 
\usepackage{subcaption}
\title{The Eccentric Disk Model for Superhumps}

\author*[a]{Stephen H. Lubow}

\affiliation[a]{Space Telescope Science Institute,\\
 3700 San Martin Drive, Baltimore, MD 21218, USA}

\emailAdd{lubow@stsci.edu}

\abstract{An important goal of the disk instability model is to explain the
superhump phenomenon. Superhumps are features found in the
light curves of binary systems, characterized by a period slightly different from the binary orbital period. In cases where the superhump
period is longer than the orbital period (positive superhumps), they
have been  interpreted as arising from an eccentric, precessing disk. This paper reviews the theory and simulations that indicate 
that the disk's eccentricity originates from a dynamical instability 
at the 3:1 resonance. The instability is described by a mode-coupling process involving the interaction of the disk eccentricity with the binary tidal potential. This instability provides critical constraints on
the nature of the disk turbulence that enables the disk to reach this
resonance. 
}

\FullConference{87th Fujihara Seminar: The 50th Anniversary Workshop of the Disk Instability Model in Compact Binary Stars (DIM50TH2025)\\
22-26 September 2025\\
Tomakomai, Japan\\}

\begin{document}
\maketitle

\section{Introduction}

Dwarf novae are highly evolved binary systems that  undergo occasional increases in brightness known as outbursts.
In these systems, a red dwarf secondary star  fills its Roche lobe and transfers mass  to a white dwarf primary star through a narrow gas stream. The stream results in the formation of a disk
around the primary.
Dwarf novae undergo outbursts of 2-3 mag over a duration of
a few days that repeat on timescales of weeks to years. 
The disk instability model (DIM) was developed to explain dwarf nova outbursts as being due to a sudden accretion event in the disk.
It originated in work by Osaki \cite{Osaki1974} and Smak \cite{Smak1976} (see review by Osaki in this conference).  
 The so-called normal outbursts are described to result from a thermal instability involving the ionization of hydrogen
based on early work by H{\={o}}shi \cite{Hoshi1979} and Meyer \& Meyer-Hofmeister \cite{Meyer1981} (see also paper by Hameury in this conference).

Some systems undergo superoutbursts that are somewhat brighter, 3-5 mag, last somewhat longer, a few weeks,
and repeat on timescales of weeks to years. 
Nearly all of these systems that undergo superoutbursts also undergo normal outbursts.
A distinctive property of a superoutburst is that it exhibits a feature in its light curve that has a period that is
slightly different from the binary orbital period  \cite{Vogt1974,Warner1975}. In some cases, the period of the feature is longer than the binary
orbital period (positive superhumps), while in other cases the feature has a shorter period (negative superhumps). This review
discusses only the positive superhumps.  Superhumps in outbursting systems are generally found during superoutbursts.
Superhumps are also found as a long-term feature, called permanent superhumps, in some nonoutbursting systems.
There have been many observations of superhump systems (e.g., \cite{Patterson1995, Kato2013}).  A review of superhump observations by Nogami \& Kato is provided in this conference.

Based on photometry and spectroscopy of dwarf nova Z Chamaeleontis, Vogt \cite{Vogt1982} proposed a model
for superhumps in which the outer part of the disk is eccentric. In this model, the difference between the superhump period and the binary orbital period is
due to the slow prograde apsidal precession of the disk, resulting in a disk beat period that is slightly longer than the binary
orbital period. The superhump radiation was modeled as being due the the stream-disk impact.
Osaki \cite{Osaki1985} computed the apsidal precession 
rate of a disk due to the gravitational effects of the secondary  by calculating the precession rate of a particle at two different estimates for the disk outer radii.  The resulting prograde precession rates of a few percent of the binary angular frequency  implied superhump periods that covered the range of observed periods for the known superhump systems. This finding provided early support for the eccentric disk model.
There are alternative models for superhumps, such as that of Smak \cite{Smak2010} 
in which the superhump radiation is provided by the impact of the gas stream with  a precessing disk of azimuthally varying thickness.

Whitehurst \cite{Whitehurst1988} later discovered in 2D Lagrangian (particle) simulations of a dwarf nova disk that disk eccentricity does in fact naturally develop. It occurs as a result of the tidal interaction
of the disk with the secondary star for binary mass ratios $q < 0.25$, defined as the secondary mass divided by the primary mass. Such mass ratios are common in superhump systems.
In these simulations a disk is formed from a burst of accretion through a mass transfer gas stream. The disk is initially compact and nearly circular with its outer edge near the so-called circularization radius of the gas stream
\cite{Lubow1975}.  It later expands due to viscosity, becomes eccentric, and precesses.
As a result of the apsidal precession, each successive tidal interaction of the secondary with the apoastron region of the disk occurs with a period
slightly longer than the binary orbital period. Each such interaction gives rise to enhanced disk dissipation  and thus provides superhump radiation.
Hirose \& Osaki \citep{Hirose1990} later confirmed the eccentric disk instability in independent Lagrangian simulations. They also suggested that the 3:1 resonance plays a role based on the behavior of ballistic particles (see also \cite{Whitehurst1989}).

The disk instability model was extended by Osaki \cite{Osaki1989} to describe superoutbursts 
by incorporating the effects of the tidal instability into the tidal-thermal instability.
Simulations have been carried out by various groups that provide support for the tidal-thermal instability disk instability model (e.g., \cite{Murray1998,Jordan2024}).
A dynamical model for the eccentric disk instability 
was provided by Lubow \cite{Lubow1991a}.
This analysis showed that an initially small disk eccentricity grows exponentially in time through a mode-coupling process. In this process, the instability results from the interaction of the eccentric disk mode with a tidal disk mode  at the 3:1 resonance.

The outline of this paper is as follows. Section \ref{sec:modecoupling} contains a review of the mode-coupling model for the eccentric disk instability.
Section \ref{sec:brdisk} discusses its application to a broad disk.
Section \ref{sec:grid} describes results of some 2D grid based simulations that applied the alpha disk model for turbulence  \cite{Shakura1973}.  
Section \ref{sec:3D} discusses possible 3D effects.
Section \ref{sec:MHD} discusses recent MHD simulations in which the disk turbulence occurs through the magneto-rotational
instability (MRI)  of Balbus \& Hawley \cite{Balbus1991}. Section \ref{sec:alpha} discusses the value of turbulent viscosity parameter alpha. Section \ref{sec:precession} discusses disk precession. Section \ref{sec:summary} contains the summary.

\section{The Mode-coupling Model for the Eccentric Disk Instability}
\label{sec:modecoupling}

\subsection{Background}
The binary is on a circular orbit. How does an initially noneccentric ring/disk
become eccentric? The disk is initially noncircular
owing to the tidal effects of the secondary, but that is not what is meant
by an eccentric disk. An eccentric disk is one that is lopsided relative to the central star
and this lopsided component of the density distribution is nearly stationary
in the inertial frame. Tidal distortions of the disk are stationary in the corotating frame of the binary,
while eccentric distortions are not. The eccentric motions are described by 
noncircular velocities of particles in the classic Kepler two-body problem but
modified by disk pressure and turbulence.

A tidally distorted disk with exactly zero initial eccentricity cannot develop eccentricity.
Eccentricity is a vector that requires a specified direction that is not defined
for a noneccentric disk. Developing eccentricity requires symmetry breaking
to impose a direction. Eccentricity can develop as an instability that
grows as a result of some small initial eccentric perturbation. 
Since such perturbations are expected to be very small, the instability
needs to grow rapidly, likely exponentially in time with a substantial growth rate, 
to produce a significant eccentricity. 

 As discussed in the Introduction, the eccentric disk instability originates
 at the 3:1 resonance, that is approximately at the radius $r$ from the primary where
 $\Omega(r)= 3 \Omega_{\rm b}$
 for Keplerian angular rotation rate $\Omega(r)$ and binary rotation rate $\Omega_{\rm b}$.
 Disk streamlines can be approximately represented as simple periodic orbits about the
 primary star \cite{Paczynski1977}. At small radii, these orbits are circular and nested.
 At larger radii, these orbits become tidally distorted and mutually cross each other.
 The minimum radius of orbit crossings is associated with the  disk outer radius,
 since strong tidal effects would occur there \cite{Papaloizou1977}. 
 This radius depends on the binary mass ratio. 
 For $q > 0.25$, the disk would be truncated interior to the 3:1 resonance and therefore
 the disk would not be affected by this resonance.
 For binary mass ratio $q < 0.25$
 the maximum disk radius lies at a larger radius than the 3:1 resonance. Therefore,
 the disk could become eccentrically unstable at small mass ratio.
  
 The description of disk streamlines as simple periodic orbits is an approximation.
 The disk experiences pressure and viscous forces that modify this description  in two ways.
 First, the disk truncation radius depends on the kinematic viscosity in the disk.
 Tidal torques acting on the disk act to cause the disk to lose angular momentum and contract,
 while viscous torques cause the outer parts of the disk to expand. The disk
 radius is determined by the condition in which these two torques are in balance
 (e.g., \citep{Goldreich1980, Artymowicz1994}).
 Second, the disk resonance has some nonzero radial width that depends
 on the sound speed and viscosity \cite{MeyerVernet1987}. So a disk need not
 exactly reach the resonance in order to experience its effects. In addition, the outer region
 of the resonance might not lie within the disk, resulting in an overall weakening of the effects of the resonance.
 
 \subsection{Eccentricity Instability Cycle}
 \label{sec:instabcyc}
 
The early simulations showed that the disk could develop eccentricity as a consequence 
of the tidal effects of the secondary star. A description of how
that eccentricity develops and the role of the 3:1 resonance was provided
by the mode-coupling model of Lubow \cite{Lubow1991a, Lubow1991b}.
This section provides an outline of that model.

The mode-coupling model describes the instability in terms of disk modes that we define
here. Consider a polar coordinate system $(r, \phi)$ centered on the primary. 
Any physical quantity in the disk can be expanded in a Fourier series
in azimuthal angle $\phi$ and time $t$ because the disk is periodic in those quantities.
A mode is defined as a component in such a Fourier series. It is labeled 
by  mode numbers  $(m, \ell)$ and is of the form
\begin{equation}
X_{m, \ell} (r, \phi, t) = Re[X_{m, \ell}(r) \exp{[i (m \phi - \ell \Omega_{\rm b}) t]]},
\end{equation}
 where $X$ is some physical quantity such as a velocity component or density and $X_{m, \ell}(r)$
 is a complex function of radius.
 The eccentric disk mode has mode numbers $(1, 0)$, ignoring the small
 frequency shift due to precession. (Precession can be included,
 but its effects are small.)
 We also consider the tidal mode $(3, 3)$ that plays an important role in the instability.
 
  \begin{figure}
  \center
   \includegraphics[width=0.4 \textwidth]{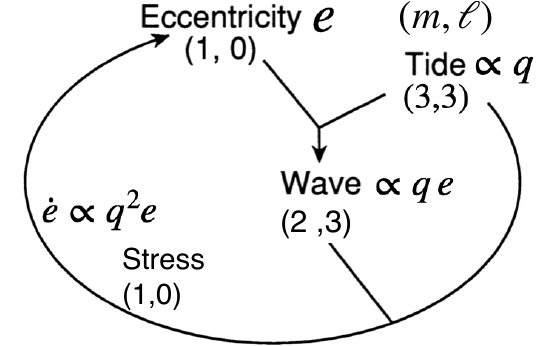} 
    \caption{Diagram of the eccentricity instability cycle. The eccentric and tidal modes interact to generate a spiral wave at the 3:1 resonance that
    in turn interacts again with the   tidal field to generate a stress that amplifies the eccentricity. }   
    \label{fig:cycle}
\end{figure}

  The interaction of two modes gives rise to an effective forcing that involves
 the product of the modes. Since both modes are periodic functions of $\phi$ and $t$, the effective forcing is also periodic. It has two 
 sets of mode numbers: one associated with the sum of the mode numbers
 and another with their difference. In this case, that means the forcing
 involves terms with $(m, \ell) = (3,3) \pm (1, 0) = (2, 3) \, {\rm and } \, (4, 3)$. Forcing in the form of
 $(2, 3)$ gives rise to a launched two arm trailing spiral wave at the eccentric
 inner Lindblad resonance \cite{Goldreich1980}. In this case, the resonance is the 3:1 resonance.\footnote{In principle,  eccentric corotational resonances could also play a role, in addition to the lowest order eccentric Lindblad resonance described here. However, interior to the 2:1
 resonance there are no lowest order eccentric corotational resonances \cite{Lubow1991a}.}
 The amplitude of this wave
 scales with the forcing that results from the interaction of the eccentricity and tide. The wave amplitude then scales as 
 $q e$ for disk eccentricity $e$. In a second step, the wave again interacts with the tide
 to provide a forcing of the form $(3,3)-(2,3)=(1,0)$ that can be shown to provide a stress
 that increases the amplitude of the eccentric mode.
 The rate of change of eccentricity in time then scales as the amplitude of the wave times the 
 amplitude of the tidal forcing, $\dot e \propto e q^2$. That is, since eccentricity
 is applied once in this two step cycle, its rate of change is linear in $e$ and therefore it
 grows exponentially in time. Because the tidal potential is invoked twice, once for each
 step in the instability cycle, the growth rate is quadratic in binary mass ratio.
 
 A detailed calculation  \cite{Lubow1991a}  in the limit of $q \ll 1$   for a ring of uniform eccentricity shows that the eccentricity
 growth rate $\lambda$ is positive  and is given by
 \begin{eqnarray}
 \lambda &=& 2.08  \, q^2 \left(\frac{r_{\rm res}}{\Delta r}\right)  \Omega_{\rm b}  
   =1.00 \, q^2 \left(\frac{  a_{\rm b}}{\Delta r} \right) \Omega_{\rm b} ,
  \label{lambda} 
 \end{eqnarray}
 where $\Delta r$ is the radial width of the ring, $a_{\rm b}$ is the binary orbital radius,
 and $r_{\rm res}= 3^{-2/3} a_{\rm b} \simeq 0.48 a_{\rm b}$  is the radius of the 3:1 resonance.
 The inverse dependence on $\Delta r$ is a consequence of the fact that
 the torque due to the launched wave is proportional to
 disk surface density at the resonance, while the resulting eccentricity increase is distributed
 over the entire mass of the ring. 
 Ogilvie \cite{Ogilvie2007} later found agreement with Equation (\ref{lambda})  in a Hamiltonian model 
 for test particles that is augmented with dissipation to resolve the resonant singularity.

 More generally, eccentricity by this mechanism can be generated by tidal
 field components with other mode numbers $(m, m)$, as occurs in planetary rings
 that are disturbed by shepherding satellites \cite{Goldreich1981}.
The lowest order eccentric Lindblad resonances that  can result in exponential
 eccentricity growth occur where
\begin{equation}
\Omega(r) = \frac{m}{m \pm 2} \Omega_{\rm b},
\end{equation}
where the plus sign 
applies to circumbinary disks and the minus sign applies to circumstellar disks. We are interested here in the latter case.
For $m=3$, we have the 3:1 resonance, as discussed above.
 For $m=2$, the resonance occurs at $r=0$ where the tidal forcing from the companion is zero; 
$m=1$ involves a retrograde disk. 
{\it The conclusion is that the 3:1 resonance is the innermost resonance that allows
eccentricity to grow exponentially in time.}  There are resonances 
at locations closer to the central star, but they are higher order and do not cause exponential growth of eccentricity. They cannot then be responsible for the instability.

The mode-coupling model makes several predictions that can be tested in numerical
simulations. Video \url{https://youtu.be/7fd-0V0jJiU} shows the results of a SPH
simulation of a disk that initially has very small eccentricity, 
although tidally distorted.  The disk is subject to only the $(3,3)$ component of the tidal
potential of the companion that is responsible for the eccentric instability according
to the mode-coupling model. The video shows that the disk develops two trailing
spiral arms, as predicted, since the spiral wave has $m=2$. In addition, the spiral arms
are predicted to rotate at angular speed $\ell \Omega_{\rm b}/m= 3 \Omega_{\rm b}/2$.
This rotation rate lies between the angular speed of the companion and the angular
speed of material at the 3:1 resonance, as is seen in the video.
At early times, there is a rotating triangular response of the disk as would be expected from a simple linear response to the $(3,3)$ tidal potential.
But at later times the primary response is the two arm spiral mode (2,3) along with the eccentric mode (1,0) as a result of the instability.

\section{Eccentricity Growth in a Broad disk}
\label{sec:brdisk}

Equation (\ref{lambda}) for the growth rate applies to a ring of uniform eccentricity.
In a broad disk, the eccentricity is expected to vary in radius.
Goodchild \& Ogilvie \cite{Goodchild2006}  applied this growth rate to a broad 2D disk
by including it as a term in an eccentricity evolution equation.
Eccentricity is a vector in Cartesian coordinates $(e_x, e_y)$ that is represented
as a complex quantity $E(r,t) = e_x + i \, e_y$.
The evolution equation for eccentricity takes the form
\begin{equation}
J(r) \partial_t E = i \partial_r(a(r) \partial_r E))+i b(r) E + J(r) s(r) E,
\label{eevolve}
\end{equation}
where $J(r) =2 \pi \Sigma(r) r^3 \Omega(r)$ is the angular momentum per unit radius in the disk,
which has unperturbed surface density $\Sigma(r)$. Quantity $b$ is  a real function of radius, while $a$ is real in the absence of dissipation, and $s$ is a complex function.
The first term on the RHS is a radial wave propagation term that depends on the disk sound speed and
the second term is a precession term due to the effects of the gravity of the secondary star and gas pressure.
The third term is the resonant term. Its real part is an
eccentric instability term based on Equation (\ref{lambda}) and the imaginary part is the resonant precession term\footnote{ Goodchild \& Ogilvie \cite{Goodchild2006} did not include the resonant precession term that was later provided  by Ogilvie \cite{Ogilvie2007} (see also \cite{Lubow1992a}). We include it in a mode calculation for the first time here.} with
\begin{equation}
s(r) = 2.98  \, q^2 \frac {w - i \omega(r)}{\omega(r)^2 +w^2}  \Omega_{\rm b}^2
\label{s}
\end{equation}
for
\begin{equation}
\omega(r) = \Omega(r) - 3  \Omega_{\rm b},
\label{om}
\end{equation}
where $w$ is the resonance frequency width \cite{Ogilvie2007}.  The local resonance growth rate $Re[s(r)]$ peaks at the resonance  and $w$ is determined
by the disk sound speed and viscosity \cite{MeyerVernet1987}.  For a narrow ring of radial extent $\Delta r$ with uniform eccentricity 
centered on a narrow resonance of frequency width $w \ll \Delta r \Omega_{\rm b}/ r_{\rm res}$,
the growth rate determined by Equation (\ref{eevolve})  is $\lambda=\int_{r_{\rm res} - \Delta r/2} ^{r_{\rm res} + \Delta r/2} Re[s(r)] dr /\Delta r$ that equals
the growth rate given by Equation (\ref{lambda}).

Equation (\ref{eevolve}) can be solved as either an evolution equation in time starting from some initial condition
or as an eigenvalue equation in which $\partial_t E(r, t) = \lambda E(r)$. The real and imaginary
parts of $\lambda$ are the growth and apsidal precession rates, respectively.
 The solutions are subject to boundary conditions.  They are typically taken to be that
$\partial_r E=0$ at the inner and outer disk radii that correspond to requiring that the Lagrangian density perturbations
vanish at the boundaries. In some cases, it may be appropriate to require $E=0$ at the disk inner radius.

\begin{figure}[htbp]
  \centering
  \begin{subfigure}{0.45\textwidth}
    \centering
    \includegraphics[width=\linewidth]{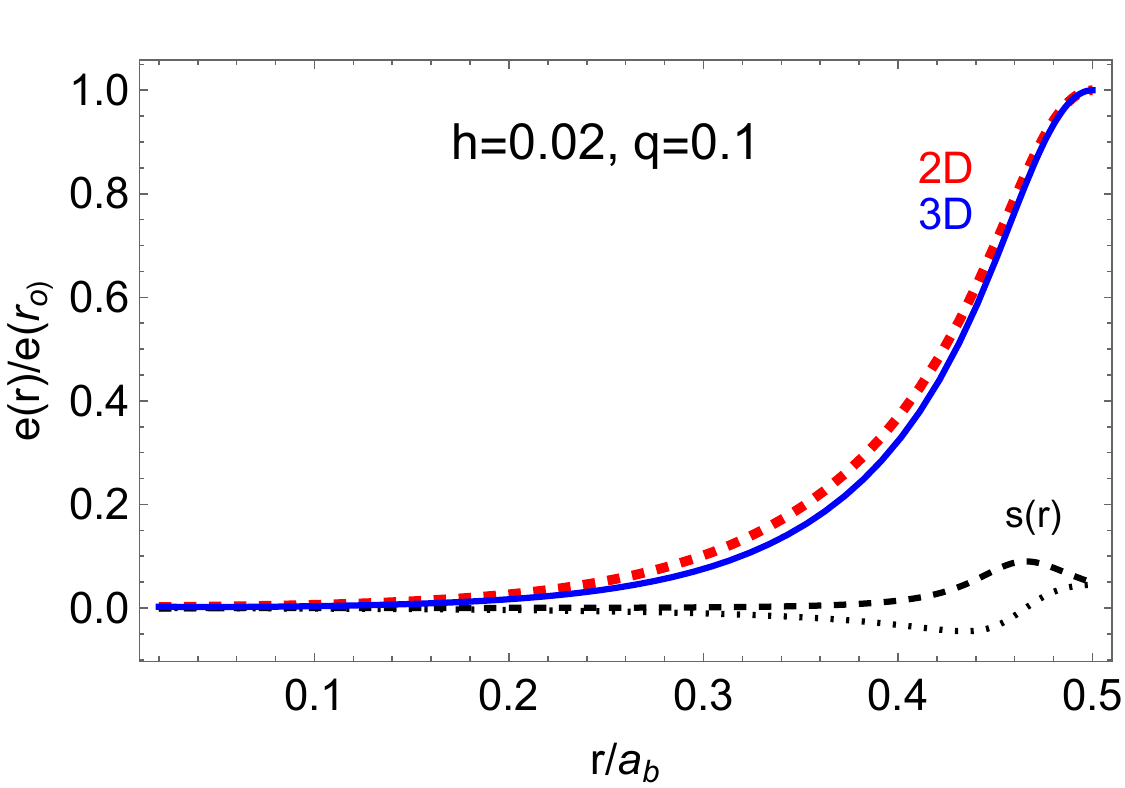}
  \end{subfigure}
  \hfill
  \begin{subfigure}{0.45\textwidth}
    \centering
    \includegraphics[width=\linewidth]{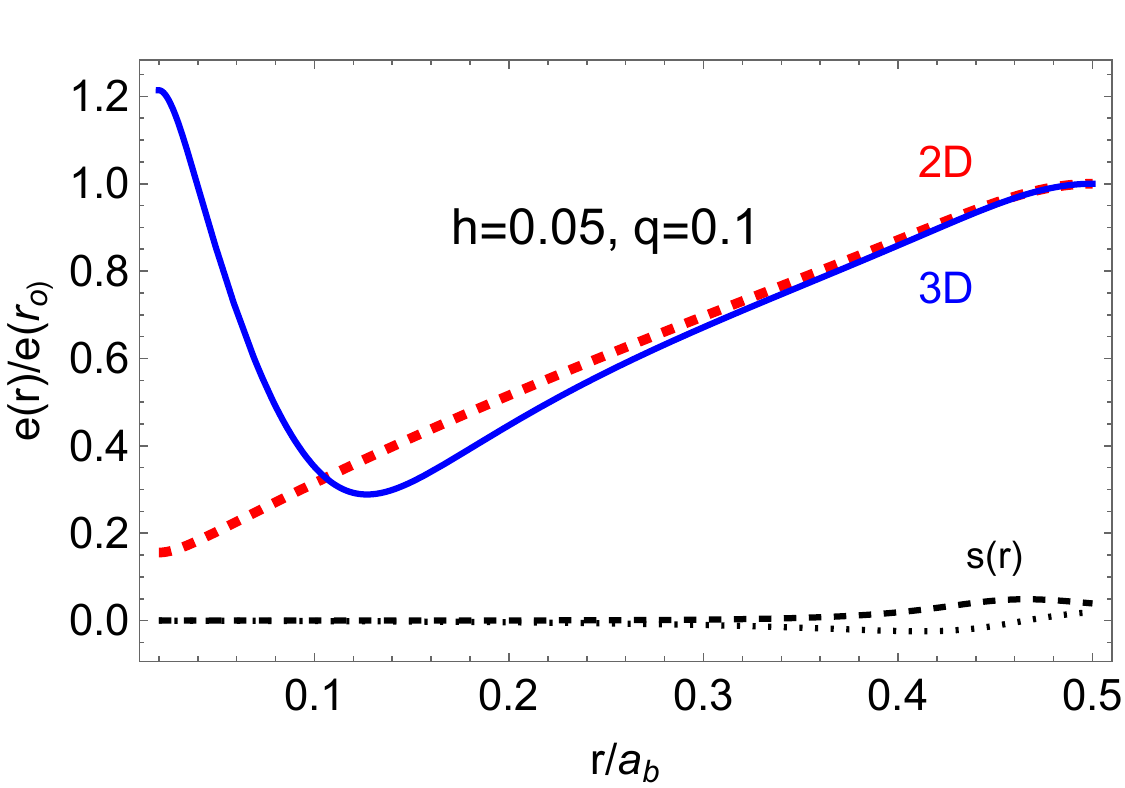}
  \end{subfigure}
    \caption{Eccentricity $e=|E|$ of the fastest growing eigenmodes that are normalized by their values at the disk outer edge $e_{\rm o}$
    plotted as a function of dimensionless radius $r/a_{\rm b}$. The binary mass ratio is $q=0.1$ in all cases. The red  dashed lines
    are for 2D disks, while the solid blue lines are for 3D disks. The black dashed and dotted lines plot the resonant growth and precession terms, respectively, given by the real and imaginary parts of Equation (\ref{s}) in units of $\Omega_{\rm b}$. The left panel is for a disk with aspect ratio $h=0.02$, similar to what is expected during a  superoutburst, while the right panel is for a warmer disk with aspect ratio $h=0.05$.}   
    \label{fig:eigenmodes}
\end{figure}

The physical solution corresponds to the fastest growing eigenmode that is the eigenmode $E(r)$ for which $Re[\lambda]$ is largest.
This typically corresponds to the fundamental eccentric mode which usually grows
much faster than other eccentric modes \cite{Lubow2010}. Figure \ref{fig:eigenmodes} plots the fastest growing  eigenmodes for disks in 2D
and 3D based on the adiabatic equations given in Teyssandier \& Ogilvie \cite{Teyssandier2016} for a disk with $\Sigma(r) \propto 1/\sqrt{r}$ and boundary conditions 
$\partial_r E=0$ at the disk inner and outer radii.  The 3D model allows for vertical motions in the disk that are a consequence of the eccentric motions as will
be discussed in Section \ref{sec:3D}.
 The disk is taken to extend slightly
beyond the 3:1 resonance. In the case of the cooler disk plotted in the left panel, the eccentricity is confined to the outer parts of the disk. This case is similar to what is expected during a  superoutburst.
The eccentricity peak is at the disk outer edge
and there is not much difference between the 2D and 3D eccentricity distributions. In the warmer disk case plotted in the right panel, 
the 2D and 3D eccentricity distributions are much broader than in the cooler disk case. 
The 3D case has a concentration of
eccentricity in the inner region of the disk. The radial width of the resonance term $s(r)$ in the warmer case is broader than in the cooler disk case because of its
higher sound speed. For the cooler disk, $\lambda = (0.061 + 0.016 i) \Omega_{\rm b}$ in the 2D case and 
$\lambda = (0.063 + 0.021 i) \Omega_{\rm b}$ in the 3D case. 
For the warmer disk,  $\lambda = (0.024 +   0.001 i) \Omega_{\rm b}$ in the 2D case and 
$\lambda = (0.022 +  0.014  i) \Omega_{\rm b}$ in the 3D case.

 These $\lambda$ values indicate eccentricity growth e-folding times that are roughly 3 to 8 binary orbital periods.
The  cooler disk case can be roughly approximated as a narrow ring of uniform eccentricity
with $\Delta r \sim 0.1 a_{\rm b}$ that results in  a growth rate $\sim 0.1 \Omega_{\rm b}$ in Equation (\ref{lambda}),  consistent with the eigenvalues. The warmer disk has a broader eccentricity distribution
and would be expected to have a lower growth rate by Equation (\ref{lambda}) due to the larger effective $\Delta r$. In addition, there is a smaller resonance overlap with the disk.
 The precession is prograde in all four cases.  The precession rate for the cooler 3D disk is in the range found in superhump observations, a few percent.

\section{Grid Based Simulations}
\label{sec:grid}

The early simulations that showed the eccentric disk instability were all carried
out with particle (Lagrangian) codes, such as SPH  
(e.g., \cite{Whitehurst1988, Hirose1990, Lubow1991b, Murray1998}).
Such codes are very versatile, but their resolution is determined by the particle density.
The resolution can be problematic in the lower density outer regions of the disk that are typically most eccentric.
The outer regions are subject to stronger effects of artificial viscosity and consequently
the disk viscosity cannot be easily controlled.
Grid based (Eulerian) codes do not suffer from this limitation because the resolution is
determined by the grid spacing that can be controlled. Several attempts were made to
find this instability in grid based codes but failed to find it (e.g., \cite{Heemskerk1994, Stehle1999}). 
The instability was found when
only the $(3,3)$ tidal field was present, as in the video of Section \ref{sec:instabcyc}.
But for the full tidal potential, no instability was found because the disk was not able to reach the 3:1 resonance. 
In some studies, no explicit viscosity was
included in the code which would prevent the disk from expanding outward, but in other studies viscosity
was included. These results cast doubt on the viability of the eccentricity disk instability
as an explanation for superhumps.

The first grid based code study to find the instability was reported by Kley, Papaloizou, and Ogilvie \cite{Kley2008}. It was published 20 years after Whitehurst's paper! This paper confirmed the eccentric disk instability and found agreement with
several predictions of the mode-coupling model. The eccentricity growth rate and final eccentricity were found to be sensitive
to the disk viscosity. For disk turbulent viscosity parameter $\alpha$ of a few percent or higher and typical
disk aspect ratios,
there was substantial eccentricity growth. 

The mode-coupling model predicts that the (2,3) disk mode plays a critical role in causing the instability.
Kley et al. found that it does grow in time. However, other modes that result from interactions
with the eccentric mode will also grow simply because the eccentricity grows. One example is the 
(3, 2) mode that is due to the interaction of the (1,0) eccentric mode with the (2,2) tidal mode. But in the mode-coupling model,
the (3,2) mode is not predicted to play a role in {\it causing} the instability. 
In fact, such modes are predicted to be a byproduct of the instability. 

However, it is possible to determine whether the (3,3) component of the tidal potential causes the instability by a running simulations in which that component of the tidal potential is removed from the full tidal potential. 
(Another way to understand which disk modes cause eccentricity growth is to examine their contributions to the growth rate \cite{Oyang2021, Ohana2025}.)
The prediction of the mode-coupling model is then that there would be no eccentricity growth.
Kley et al. ran this test and found that when this component of the tidal field is removed,
the instability weakens considerably.
In particular, for $q=0.1$, the eccentricity grew by a factor of roughly 100 times slower than
if the full potential were applied. But for $q=0.2$, the growth was faster but 5 times slower than
if the full potential were applied. This led them to suggest that these results could be explained by the effects of
the (1,1) tidal field interacting with the effects of the (2,2) tidal field  to produce (3, 3) forcing. 
The (3, 3) forcing is still operating as in Figure \ref{fig:cycle}, but in this case it is generated by the interaction of these other modes.
 The (3, 3) mode  again interacts with the eccentricity to generate the (2,3) wave still at the 3:1 resonance but at higher order in $q$.
Other mode-couplings could  occur 
to produce (3,3) forcing.
This result suggests that the growth rate is then a power series in $q \ll 1$ for which
Equation (\ref{lambda}) is the lowest order term and is reasonably accurate for $q \le 0.2$.

\section{3D effects}
\label{sec:3D}

Most grid based simulations have been carried out in 2D.
There are some potentially important effects that can occur in 3D.
One effect is due to the fact that an eccentric disk is not in hydrostatic equilibrium.
As a fluid element undergoes radial motions due to its eccentricity, it experiences
a time-varying vertical gravity due to the central star. This gives rise to vertical motions
in the disk that cause its thickness to vary along streamlines as found by Ogilvie \& Barker \cite{Ogilvie2014}.
The result is a precessing disk of varying thickness. Its structure is similar to that 
postulated by Smak \cite{Smak2010} as part of a model for superhumps.

Another 3D effect is on disk precession. Ogilvie \cite{Ogilvie2008}  found that precession due to gas
pressure operates somewhat differently in a 3D disk compared to a 2D disk because of
the departures from hydrostatic balance. This typically reduces the retrograde effects of pressure.
We have seen some of these effects at work in Section \ref{sec:brdisk}. The precession
rates of the 3D disks are larger than in the corresponding 2D disks plotted in Figure \ref{fig:eigenmodes}
since the retrograde effects of pressure are weaker in 3D.
The often negative precession rates found in Kley et al. may have been partly a consequence of
their use of a 2D code.

\section{MHD Simulations}
\label{sec:MHD}

Prior to 2021 all superhump simulations had adopted the alpha disk model \cite{Shakura1973} to describe the effects of disk turbulence. This model 
is limited by its simplified treatment of the turbulent viscosity. It is based
on a scaling argument that the turbulent eddies are subsonic and limited
in size to the disk thickness. The kinematic viscosity is then taken 
to be $\nu = \alpha c_{\rm s} H$ where $c_{\rm s}$ is the gas sound speed
and $H$ is the disk thickness. Parameter  $\alpha < 1$ 
can be a function of radius and time. 
This prescription has the virtue that the viscosity is easy to calculate and captures
some essential physics. On the other hand, $\alpha$ is not known from first principles because the instability
that gives rise to the turbulence is unspecified. 

Disks with magnetic fields can be subject to the magneto-rotational instability  (MRI) \cite{Balbus1991} that results in disk turbulence. In principle, turbulent MHD disks can be modeled
more accurately as a product of the MRI than with the alpha disk model
(see paper by Hirose in this conference). There is, however, a high computational
cost of including the magnetic field in disk simulations.  Many of the MRI simulations
are carried out in a box that represents a small region of the disk (e.g., \cite{Hirose2014}). 
Simulating superhumps is better done in a global simulation that demonstrates the effects
of the 3:1 resonance and the distribution of eccentricity in radius.

Oyang, Jiang, \& Blaes 
\cite{Oyang2021} carried out the first MHD  
superhump disk simulation. They conducted a  3D global simulation involving a binary with mass ratio $q=0.1$ for case of permanent 
superhumps in which the disk temperature distribution in radius is fixed in time. They  compared these
results to those of 2D alpha disk simulations that used a grid based code. 
The initial configuration of the MHD simulation had no disk. The disk was formed
from a gas stream that carried magnetic loops with no net vertical flux, that is no net flux perpendicular to the disk.
The resulting disk became turbulent through  the MRI. However, the disk eccentricity did not grow significantly over the
course of the simulation, about 250 binary orbital periods. 
They also attempted to increase
the level of disk turbulence by injecting more magnetic loops into the disk, but the
eccentricity still did not grow. The reason for the lack of eccentricity growth is that the
disk did not expand enough to reach the 3:1 resonance. The effective value of $\alpha \simeq 0.01$
is typical of MRI simulations that have no vertical magnetic flux. 

Their 2D alpha disk simulation showed a similar
lack of eccentricity growth for $\alpha=0.01$ and a similar lack of disk material at the 3:1 resonance. But with $\alpha=0.1$ and 0.2, the disk did reach the resonance and the eccentricity
grew exponentially in time.  These higher alpha 
simulations also showed agreement with predictions of the
mode-coupling model. In particular, the  (2, 3) wave was found to provide the largest
contribution to eccentricity growth. 
In addition, the next most important contribution is due a mode of eccentric form that may be produced by the eccentric stress in the second step of the instability cycle (Figure \ref{fig:cycle}).
 These results suggest that the eccentric disk
instability can take place if the disk is sufficiently turbulent. But they indicate that MRI without a vertical
flux cannot achieve the required level of turbulence to reach the 3:1 resonance.

More recent MHD simulations were carried out by Ohana et al. \cite{Ohana2025}
with some important modifications to the Oyang et al. model. In these MHD simulations,
  the initial 
magnetic field was vertical. The disk was also initially vertically unstratified with no vertical gravity in
order 
to prevent the development of magnetic winds. Simulations were also run for an alpha disk model for $\alpha =0.1$ using the same code otherwise. In one MHD simulation and in the alpha disk simulation,
there was no initial disk and the disk was built by the gas stream.
In this case, the MHD disk did not show significant eccentricity, while the alpha 
disk did  exhibit eccentricity growth from 100 binary orbits  to 
the end of the simulation at 140 binary orbits.
The MHD disk was again not able to reach the resonance over the time of the simulation.
The MHD disk had an average effective $\alpha > 0.1$, but the effective alpha in the outer
parts of the disk was lower.

Three other MHD simulations were run without a gas stream. In these simulations, the disk initially covered the 3:1 resonance.
In all cases the disk developed eccentricity and agreed with the predictions of the mode-coupling model. The disk continued to cover the resonance at later times.

The magnetic field plays a dual role in the evolution of disk eccentricity.
In both MHD studies, it was found that the magnetic stresses overall acted to damp 
eccentricity. On the other hand the resulting disk turbulence acts to expand the disk and
in principle allows it  to reach and continue to cover the 3:1 resonance. Based on the simulations in which the disk initially covers the resonance, it is clear that if
an MHD disk can reach the resonance, it can overcome the damping effects
of magnetic stresses and grow eccentricity by the mode-coupling mechanism.
This result is similar to what happens in alpha disks because they are similarly subject to this same
set of dual effects and can grow eccentricity. Growth occurs because the eccentricity growth rate due to the mode-coupling cycle for superhump binaries with $q \sim 0.1$
is typically greater than the eccentricity damping rate due to viscosity. 

In the MHD disk simulations that initially covered the resonance,
the disk eccentricity undergoes a series of episodes of radial disk breaking into two regions separated by a
region of circular orbits.  There is an inner eccentric region that is somewhat similar to what is found for the 3D warmer disk in Figure \ref{fig:eigenmodes} that does not involve a magnetic field. But in the MHD case, the inner eccentricity structure is likely due to effects of the magnetic field, as has been also found by Chan, Piran, \& Krolik \cite{Chan2024} in the context of tidal disruption events. 
In the binary case studied by Ohana et al., the eccentricity throughout the disk is  
ultimately provided by the 3:1 resonance. 

\section{The Value of Alpha}
\label{sec:alpha}

We have seen in the last section that the value of alpha in the outer parts of a disk plays a critical
role in determining whether a disk can reach the 3:1 resonance and become eccentric.
This sensitivity to alpha comes about because of the close competition between tidal and viscous torques
near the 3:1 resonance in typical dwarf novae with binary mass ratios $q \sim 0.1$.
The value of alpha has been estimated for dwarf novae based on the decay times of outbursts \cite{Smak1999, Kotko2012}.
Typically, these values are $\alpha \sim 0.2$. Such values are higher than found in MRI simulations \cite{King2007, Martin2019, Blaes2025} without a vertical magnetic flux. Some simulations have shown that the MRI turbulence can be enhanced
by convection to reach these higher values \cite{Hirose2014, Coleman2016}.  The convection arises at the low temperature range of the so-called hot branch in the disk instability model where hydrogen becomes
partially ionized. The enhanced turbulence however covers a small range of the hot branch. 
Consequently, this mechanism likely does not apply to disks in permanent superhump binaries that are expected to be well above the temperature range for convection in the DIM.

In the case of superhump binaries, a vertical magnetic flux was not sufficient to allow a disk formed by a gas stream to reach the 3:1
resonance \cite{Ohana2025}.  In the alpha disk case with a gas stream, the eccentricity was developed towards the end of the simulation.
It is possible that the MHD simulation would have shown eccentricity growth if run to longer times. 
Another possibility, suggested by Nixon, Pringle, \& Pringle  \cite{Nixon2024}, is that current MHD simulations have inadequate resolution, resulting in significant artificial magnetic diffusivity. The diffusivity limits the  magnetic field strengths that can be achieved by disk shearing, thereby producing lower levels of disk turbulence. They suggest that the current codes have diffusivities that are too high by several orders of magnitude than would be needed to reach the alpha values indicated by dwarf nova outbursts.

\section{Precession}
\label{sec:precession}

Precession rates provide a connection between observation and theory of superhump disks.
The eccentricity distributions and precession rates of the  
eccentric modes in 2D disks were determined by Hirose \& Osaki \cite{Hirose1993} and later by Goodchild \& Ogilvie \cite{Goodchild2006} who included the effects of the eccentricity growth.
They showed that the overall precession rate of a disk
is determined by a  mean of the local precession rates at each radius that are weighted by the angular momentum per unit radius times the square of the local eccentricity.
The overall apsidal precession rate of a disk depends on several factors that are listed below.
 \begin{itemize}
\item Gravitational forcing by the companion first examined by Osaki \cite{Osaki1985} is often the largest contribution and provides a prograde precession rate contribution. 
\item Pressure provides a retrograde contribution \citep{Lubow1992a, Murray2000, Goodchild2006, Teyssandier2016}.  2D and 3D disks give rise to different pressure induced precession rates, as discussed
in Section \ref{sec:3D}.
\item Resonant stresses can contribute \cite{Lubow1992a, Ogilvie2007}. They were found in simulations to provide a prograde
contribution when the eccentricity is growing exponentially.
\item Magnetic fields can contribute,  but there are limited results thus far. A simulation by
Ohana et al. \citep{Ohana2025} showed that magnetic fields caused retrograde precession  at later times.
\item The gas stream provided a prograde contribution to precession in the simulations by Kley et al. \cite{Kley2008}.
\end{itemize}

The various precession contributions have both a direct and indirect influence on the overall precession rate.
The indirect influence comes about because they each can affect the eccentricity distribution in the disk that in
turn modifies the weightings of all contributions in radius. In addition, each precessional contribution also depends
on the disk aspect ratio because it modifies the eccentricity distribution (see  Figure \ref{fig:eigenmodes} and \cite{Lubow2010}). 

The precession rates are observed to vary in time 
in three identified stages (A, B, and C) \cite{Kato2009}. The superhump period is shorter in the later stage C than in the earlier stage A. 

\section{Summary}
\label{sec:summary}

The results of theory and simulations show that the 3:1 resonance is 
responsible for the eccentric disk instability. Both particle based and grid based alpha disk
simulations \citep{Kley2008, Murray1998, Oyang2021, Ohana2025} 
agree well with the predictions of the mode-coupling model for the eccentric disk instability 
(\cite{Lubow1991a}; see Section \ref{sec:instabcyc}), provided  that the turbulent viscosity parameter alpha is several percent or greater. 
Such alpha values are beyond the range typically achieved in MRI simulations of disks without a net vertical
magnetic flux \cite{King2007, Martin2019, Blaes2025}.
MHD disks  can also
grow eccentricity in a way that also agrees well with the mode-coupling model, if the disk
initially has a vertical magnetic flux and initially covers the 3:1 resonance \cite{Ohana2025}. However, MHD simulations
that build the disk from the mass transfer gas stream
have not yet been able to show eccentricity growth because the disk was not able to reach
the 3:1 resonance \cite{Oyang2021, Ohana2025}.
  
Disk precession provides an important connection between theory and observations.
The precession rate is typically dominated by the prograde effects of the gravitational forces of the companion.
However, precession is a delicate process that depends on other effects as well (see Section \ref{sec:precession}). 
Consequently, we do not expect an exact relation between the superhump period normalized by the binary
orbital period and the binary mass ratio.
 
 In superhump binaries, the disk's ability to reach the 3:1 resonance and be subject to the mode-coupling instability depends on
 the close competition between tidal forces that act to contract the disk and viscous forces that act to expand
 the disk. As a result, the instability provides a sensitive probe of the conditions in the outer region 
 of a disk. This probe is particularly important in the study of MHD disks where the MRI provides the turbulence and effective
 viscosity.  However, there are computational challenges in carrying out such simulations (see Section \ref{sec:alpha}). Future studies hold the potential to provide key insights into the magnetic field strengths and topologies. Important progress has already been made \cite{Oyang2021, Ohana2025}.  
 

\bibliographystyle{JHEP}
\bibliography{ref}

\end{document}